\documentstyle[preprint,aps,eqsecnum]{revtex}

\def\simlt{\mathrel{\lower2.5pt\vbox{\lineskip=0pt\baselineskip=0pt
           \hbox{$<$}\hbox{$\sim$}}}}
\def\simgt{\mathrel{\lower2.5pt\vbox{\lineskip=0pt\baselineskip=0pt
           \hbox{$>$}\hbox{$\sim$}}}}

\begin{document}

\title{Space-Time, General Covariance, Dirac-Bergmann Observables and
Non-Inertial Frames.}

\author{{Luca Lusanna}}

\address{Sezione INFN di Firenze, Polo Scientifico, via Sansone 1,
50019 Sesto Fiorentino (Firenze), Italy\\ E-mail:
lusanna@fi.infn.it}

\maketitle

 \hfill

Talk given at the 25th Johns Hopkins Workshop {\it 2001: A
Relativistic Spacetime Odyssey}, Firenze 3-5 September 2001.

\hfill\hfill

Both particle physics and all the approaches to gravity make use
of variational principles employing singular Lagrangians \cite{1}.
As a consequence the Euler-Lagrange equations cannot be put in
normal form, some of them may be non independent equations (due to
the contracted Bianchi identities) and a subset of the original
configuration variables are left completely or partially
indetermined. Moreover, even with reasonable boundary conditions
at spatial infinity \footnote{We consider only non-compact
spacetimes both in the special and general relativistic context,
since a general relativistic description of particle physics must
reduce to a field theory on Minkowski spacetime when the Newton
constant $G$ is switched off. }, the fact that often
Euler-Lagrange equations, like in the case of Einstein's
equations, are not a hyperbolic system of partial differential
equations implies a very difficult and nearly untractable initial
value problem in configuration space.

This state of affairs is a source of an endless number of problems
at the ontological level for philosophers of science. Instead in
physics we accept the fact that the historical development of
particle theory and of general relativity has selected certain
variational principles and certain configuration variables as the
most convenient to englobe useful properties like locality,
manifest Lorentz covariance, minimal couplings (the gauge
principle), general covariance. However all this leads to the
necessity of a division of the initial configuration variables of
any theory in two groups:

i) the {\it non determined gauge variables};

ii) the {\it gauge-invariant observables} with a deterministic
evolution.

But this process is in conflict with locality, manifest Lorentz
covariance, general covariance and, moreover, the configuration
space manifestly covariant approach has no natural analytical
tools to perform this separation. Even when this is possible, like
in the radiation (or Coulomb) gauge for electromagnetism in
Minkowski spacetime, we are not yet able to quantize, regularize
and renormalize this formulation with only physical degrees of
freedom since locality is lost \footnote{There is no no-go
theorem, only the mathematical incapacity to treat the
singularities \cite{2}.}. Therefore particle physics in Minkowski
spacetime relies on the strategy of quantizing all the
configuration variables in an auxiliary unphysical Hilbert space
${\cal H}_{unphys}$ and then in making a quantum reduction to a
physical Hilbert space ${\cal H}_{phys}$, which, however, {\it in
general is not a subspace} of ${\cal H}_{unphys}$. The winning
strategy is the BRST approach \footnote{It puts control on the
infinitesimal gauge transformations but not on the finite ones. }
(with the BFV variant), but this is also the strategy of the
group-theoretical, algebraic or geometric quantization approaches.
Even if perturbative regularization and renormalization are well
defined, BRST observables are well defined algebraic quantities
and the phenomenological applications are successful, there is no
clear hint of how to solve open problems like the determination of
the {\it physical scalar product} of ${\cal H}_{phys}$ and whether
global obstructions like the Gribov ambiguity are either only
mathematical obstructions, to be eliminated with a suitable choice
of the function space for the fields, or carriers of physical
information needed for instance to explain quark confinement. The
alternative {\it first reduce, then quantize} is also very
difficult to explore, because the classical reduced theory, when
can be defined, lives usually in a topologically highly
non-trivial configuration space. Incidentally, this is also the
reason \cite{3} why the path integral approach is not well defined
globally.

Another non trivial aspect of the need of the division between
gauge variables and deterministic observables is the connection of
the latter with measurable quantities. Since, at least at the
classical level, the electromagnetic measurable quantities are the
local electric and magnetic fields, we can extrapolate that the
non-local radiation gauge observables, i.e. the transverse vector
gauge potential and the transverse electric field, are also
measurable. But in the case of the non-Abelian Yang-Mills gauge
theories for the strong and weak interactions the connection
between gauge invariant observables and measurable quantities is
still poorly understood. With trivial principal bundles, in
suitable weighted Sobolev spaces in which both the aspects of
gauge symmetries and gauge copies of the Gribov ambiguity are
absent, with all the fields tending suitably to zero in a
direction-independent way so that the non-Abelian Yang-Mills
charges are well defined, in a non-Abelian radiation gauge the
non-local observables are again the transverse Yang-Mills vector
gauge potentials and elecric fields \cite{4}. In more general
cases it is unknown and we have only the algebraic BRST
observables, which carry topological informations on the reduced
configuration space but no global description of it, since
perturbative methods cannot describe finite gauge transformations
and the problems connected with the associated infrared
divergencies \cite{5}.

When we come to general relativity in Einstein formulation these
problems become both more complex and more basic. More complex
because the Lie groups underlying the gauge groups of particle
physics are replaced by diffeomorphism groups
\footnote{Reparametrization invariant theories in Minkowski
spacetime for particles and strings and parametrized Minkowski
theories for every isolated system \cite{1,6} also have
diffeomorphism groups as gauge groups.}, whose group manifold in
large is poorly understood. More basic because now the action of
the gauge group is not in an inner space of a field theory on a
background spacetime, but is an extension to tensors over
spacetime of the diffeomorphisms of the spacetime itself. This
reflects itself in the much more singular nature of Einstein's
equations \footnote{Four of them are not independent from the
others due to the Bianchi identities, four are only restrictions
on the initial data and only two combinations of Einstein's
equations and their gradients depend on the {\it accelerations}
(the second time derivatives of the metric tensor).} with respect
to Yang-Mills equations. This fact has the dramatic consequence to
{\it destroy any physical individuality of the points of
spacetime} as evidentiated by Einstein's {\it hole argument}
\cite{7} in the years (1913-16) of the genesis of the concept of
general covariance. Only the idealization ({\it point-coincidence
argument}) according to which all possible observations reduce to
the intersections of the worldlines of observers, measuring
instruments and measured physical objects, convinced Einstein to
adopt general covariance and to abandon the physical objectivity
of spacetime coordinates. This argument, after a long oblivion,
was resurrected by Stachel \cite{8} and then by Norton \cite{9}
and others as a basic problem \cite{10} in our both ontological
\footnote{See the modern debate \cite{10} about the failure of
{\it manifold substantivalism} to reconcile the need for
determinism in classical physical laws and a {\it realistic}
interpretation of the mathematical 4-manifold describing
spacetime. } and  physical understanding of {\it spacetime} in
general relativity.

General relativity is formulated starting from a mathematical
pseudo-Riemannian 4-manifold $M$ endowed with an atlas of
coordinate charts. As a topological Hausdorff space, the points of
$M$ can be identified only by means of coordinates, i.e.
quadruples of real numbers.

Einstein's equations are trivially form-invariant under {\it
passive} diffeomorphisms, namely arbitrary changes of coordinates
in $M$, being tensorial equations. Let us now consider {\it
active} diffeomorphisms ${\cal F}: M \mapsto M$, under which a
point $P$ is sent into the point ${\cal F}(P)$. In a coordinate
chart ${\cal C}$, if $x^{\mu}(P)$ are the coordinates of $P$, then
$x^{\mu}({\cal F}(P)) = f^{\mu}(x(P))$ are the coordinates of
${\cal F}(P)$. Given a 4-metric tensor ${}^4g$ on $M$ with
components ${}^4g_{\mu\nu}(x)$ in the chart ${\cal C}$, we can
define a new metric tensor ${}^4g^{'}$ (the {\it drag-along} of
${}^4g$) by asking that in every coordinate chart containing $P$
and ${\cal F}(P)$ we have ${}^4g^{'}_{\mu\nu}(x({\cal F}(P))) =
{}^4g_{\mu\nu}(x(P))$, so that ${}^4g^{'}_{\mu\nu}(x(P)) \not=
{}^4g_{\mu\nu}(x(P))$. If ${}^4g$ is a solution of Einstein's
equations, also ${}^4g^{'}$ is a solution, namely {\it active
diffeomerphisms are dynamical symmetries} of Einstein's equations
sending solutions into solutions.

Let us consider a {\it hole} ${\cal A}$ in $M$ not containing
matter and an active diffeomorphism which is the identity outside
${\cal A}$, where all the matter resides and where suitable
initial data and boundary conditions have been chosen. Then we
have the situation that inside ${\cal A}$ we have two different
solutions ${}^4g$ and ${}^4g^{'}$ which however coincide outside
${\cal A}$. Therefore there is no deterministic propagation of the
solution from outside ${\cal A}$ to the interior of ${\cal A}$ and
no way to attribute a physical individuality to the points of
${\cal A}$ (which is the real gravitational field in a point?).
The only way to avoid this lack of determinism is to say with
Dirac that {\it only the equivalence class} of all the 4-metric
tensors solution of Einstein's equations modulo active and passive
diffeomorphisms is a {\it physical solution}.This property was
named by Earmann and Norton \cite{11} {\it Leibniz equivalence}
\footnote{See Ref.\cite{12} for the improper use of the name of
Leibniz in a theory containing fields instead of only mechanical
particles like in Newton-Leibniz times.} and points toward the
necessity of considering the 4-manifold $M$ {\it and} the 4-metric
tensor ${}^4g$ \footnote{It gives the {\it dynamical
chronometrical structure} of $M$, differently from what happens in
special relativity, where both Minkowski spacetime and the
chronometrical structure (the clocks and the rods and therefore
the whole theory of measurements) have a non-dynamical absolute
existence. This also explains because till now we do not have a
well defined theory of measurement in general relativity. As a
consequence, there is a fundamental difference between generally
covariant field theories like Einstein's general relativity and
theories on a background like string theory. If ${\cal M}$ theory
exists and if its mathematical apparatus is not too exotic so that
some form of general covariance can be defined, it too will have
to face the {\it hole argument}. } as an inseparable entity $(M,
{}^4g)$ in general relativity.

Let us remark that Bergmann and Komar \cite{13} were able to give
a passive re-interpretation of active diffeomorphisms, by showing
that Einstein's equations are form-invariant not only under
coordinate transformations $x^{{'}\, \mu} = h^{\mu}(x)$, but also
under generalized transformations of the form $x^{{'}\, \mu} =
h^{\mu}(x, {}^4g(x))$.

In physics the {\it hole argument} is considered an aspect of the
fact that also Einstein's theory is interpreted as a gauge theory.
The Leibnitz equivalence is nothing else than the selection of the
gauge invariant observables of the theory. But now, differently
from Yang-Mills theories, we have lost the physical interpretation
of the underlying mathematical 4-manifold, and this suggests that
a different interpretation of the gauge variables of generally
covariant theories with respect to Yang-Mills theories is needed.
Moreover, one would like that the observables in general
relativity be {\it Bergmann observables} \cite{14}, namely
(possibly local) quantities independent from the choice of
coordinates. For instance the measurements of matter quantities in
general relativity are idealized as performed by {\it test
timelike observers} (either isolated or belonging to a congruence)
{\it endowed with a tetrad} (the unit 4-velocity of the observer
plus gyroscopes for the spatial triad): to avoid coordinate
singularities and/or misunderstandings, the only acceptable
measurable quantities are the {\it tetradic} components of matter
tensors, which are Bergmann observables.

Let us remark that we do not accept the point of view
\footnote{See the material reference fluids in Refs.\cite{15}.}
that matter is necessary for the physical individuation of
spacetime points. The conceptual problem already exists in vacuum
general relativity and has to be solved only by using the vacuum
gravitational field, which has an ontological priority over all
matter fields since it tells them how to move causally \cite{16}.
Test matter will only enter in the actual measurement process at
an operational level.

As already said the manifestly covariant configuration space
approach has no natural tool to make a clean separation between
gauge variables and a basis of gauge invariant (hopefully
measurable) observables. Instead, at least locally, the
Hamiltonian formulation has natural tools for it, namely the
Shanmugadhasan canonical transformations \cite{17}.

The singular Lagrangians of particle physics and general
relativity imply the use of Dirac-Bergmann theory \cite{18,19,1}
of Hamiltonian constraints \footnote{Namely of presymplectic
geometry, the theory of a closed but degenerate two-form, in
geometrical language. } and only the constraint submanifold of
phase space is relevant for physics. Let us consider a
finite-dimensional system with configuration space $Q$ with global
coordinates $q^i$, $i=1,..,N$ described by a singular Lagrangian
$L(q,\dot q)$ [${\dot q}^i(\tau ) = d\, q^i(\tau ) / d\tau$]. Let
the Dirac algorithm produce the following general pattern:

i) $m < N$ first class constraints $\phi_{\alpha}(q,p) \approx 0$,
of which the first $m_1 \leq m$ are primary, with the property
that the Poisson brackets of any two of them satisfies $\{
\phi_{\alpha}(q,p), \phi_{\beta}(q,p) \} =
C_{\alpha\beta\gamma}(q,p)\, \phi_{\gamma}(q,p) \approx 0$ ;

ii) $2n$ second class constraints, corresponding to pairs of
canonical variables which can be eliminated by going to Dirac
brackets;

iii) a Dirac Hamiltonian $H_D = H_c + \sum^m_{\alpha = m_1+1}\,
r_{\alpha}(q,p)\, \phi_{\alpha}(q,p) + \sum_{\alpha =1}^{m_1}\,
\lambda_{\alpha}(\tau )\, \phi_{\alpha}(q,p)$, where the
$\lambda_{\alpha}(\tau )$'s are arbitrary functions of time, named
{\it Dirac multipliers}, associated only with the {\it primary}
first class constraints \footnote{The use of the first half of
Hamilton equations, ${\dot q}^i = \{ q^i, H_D \}$, shows that the
Dirac multipliers are those primary velocity functions
($g_{\alpha}(q,\dot q) = \lambda_{\alpha}(\tau )$ on the solutions
of Hamilton equations) not determined by the singular
Euler-Lagrange equations. It can be shown that this arbitrariness
implies that also the secondary velocity functions
$r_{\alpha}(q,p) = {\tilde r}_{\alpha}(q,\dot q)$, $\alpha =
m_1+1,..,m$, in front of the secondary (and higher) first class
constraints in $H_D$, are not determined by the Euler-Lagrange
equations. Therefore each first class constraint has either a
configuration or a generalized velocity as an arbitrary partner.}.
In phase space there will be as many arbitrary Hamiltonian gauge
variables as first class constraints: they determine a
coordinatization of the {\it gauge orbits} inside the constraint
submanifold. The first class constraints are the generators of the
Hamiltonian gauge transformations under which the theory is
invariant and a gauge orbit is an equivalence class of all those
configurations which are connected by gauge transformations
(Leibnitz equivalence). The $2(N-m-n)$-dimensional reduced phase
space is obtained by eliminating the second class constraints with
Dirac brackets and by going to the quotient with respect to the
gauge orbits, or equivalently by adding as many gauge fixing
constraints as first class ones so to obtain $2m$ second class
constraints.

At least locally on the constraint submanifold the family of
Shanmugadhasan canonical transformations $q^i, p_i \mapsto
Q^{\alpha}, P_{\alpha} \approx 0, {\bar Q}^{\beta} \approx 0,
{\bar P}_{\beta} \approx 0, Q^A, P_A$, $\alpha =1,..,m$, $\beta =
1,..,n$, allows

i) to Abelianize the first class constraints, so that locally the
constraint submanifold is identified by the vanishing of a subset
of the new momenta $P_{\alpha} \approx 0$;

ii) to identify the associated Abelianized gauge variables
$Q^{\alpha}$ as coordinates parametrizing the gauge orbits;

iii) to replace the second class constraints with pairs of
canonical variables ${\bar Q}^{\beta} \approx 0$, ${\bar
P}_{\beta} \approx 0$;

iv) to identify a canonical basis of {\it gauge invariant Dirac
observables} with a deterministic evolution determined only by the
gauge invariant canonical part $H_c$ of the Dirac Hamiltonian.

This is the tool of the Hamiltonian formalism, lacking in the
configuration space approach, which allows to make the division
between arbitrary gauge variables and deterministic gauge
invariant observables.

Since the (in general non local) Dirac observables give a
coordinatization of the classical reduced phase space, it will
depend on its topological properties whether a given system with
constraints admits a subfamily of Shanmugadhasan canonical
transformations globally defined. When this happens the system
admits {\it preferred} global separations between gauge and
observable degrees of freedom. Therefore it is important  to
understand all the topological and/or singularity properties of
the original configuration space, of the constraint submanifold
and of the reduced phase space and to study how to solve the
constraints, because otherwise there is no hope to get a global
control on the system and one has only formal formulations of the
dynamics.

In field theory the Shanmugadhasan canonical transformations are
used in a heuristic way \cite{1}, because their existence has not
yet been proved due to the fact that important constraints like
the Yang-Mills Gauss laws and the ADM supermomentum constraints
are partial differential equations of ellyptic type, which may
admit zero modes according to the choice of the function space
(see the Gribov ambiguity). In any case the identification of
canonical bases of Dirac observables is so important for
understanding the dynamics, that we can consider the assumption
that certain Shanmugadhasan canonical transformations globally
describe certain systems even in cases with non-trivial topology
as a first approximation for extracting the main non-topological
properties of a system \cite{5}.

In special relativity another source of complications originates
from the fact that the constraint submanifold of an isolated
system is the disjoint union of Poincare' strata \cite{1,6}: all
the configurations in each stratum have the conserved total
4-momentum belonging to a well defined type of Poincare' orbit.
The main stratum contains all the timelike configurations with
$P^2 > 0$. Therefore we have to define separate families of
Shanmugadhasan canonical transformations for each stratum: they
are not only adapted to the constraints but {\it also} to the
little group of the Poincare' orbit. As a consequence, after a
separate canonical reduction for each stratum, some of the Dirac
observables will no more be manifestly Lorentz covariant, but they
will only be covariant under the little group of the Poincare'
orbit.

See Ref.\cite{1} for a review of the special relativistic systems,
including the $SU(3) \times SU(2) \times U(1)$ particle standard
model, to which this type of canonical reduction has been applied
with the determination of the Dirac observables.

To get a universal control on this breaking of Lorentz covariance
and to put all special relativistic isolated systems in a form
oriented to the coupling to the gravitational field, we followed
the indications of Dirac \cite{18} to reformulate
\cite{1,6,20,21,22,23} all isolated systems on arbitrary ({\it
simultaneity and Cauchy}) spacelike hypersurfaces, leaves of the
foliation associated to a 3+1 splitting of Minkowski spacetime
\footnote{It is the classical basis of Tomonaga-Schwinger quantum
field theory, in which the classical fields acquire the non-local
information about the {\it equal-time} surfaces, information which
is absent in the standard manifestly Lorentz covariant approach.
This is the piece of information lacking in the traditional
definition of the asymptotic states of Fock space as tensor
products of free particles: in such a state an asymptotic particle
may live in the absolute future of another one and this leads to
the spurious solutions of the Bethe-Salpeter equation in the
theory of relativistic bound states. Let us also remark that for
relativistic particles this description requires the choice of the
sign of the energy of each particle. For all these topics see
Ref.\cite{1}.}. The embeddings $z^{\mu}(\tau ,\vec \sigma )$
($\tau$ and $\vec \sigma$ are coordinates adapted to the foliation
with the scalar time parameter labeling the leaves) of these
hypersurfaces $\Sigma_{\tau}$ are {\it new configuration
variables} and the induced metric on the hypersurfaces is a
function of the embeddings. If we take the Lagrangian of the
system coupled to an external gravitational field and we replace
the 4-metric with the metric induced by the embedding, we get the
singular Lagrangian for the embeddings plus the system (see
Ref.\cite{1,6} and Appendix A of Ref.\cite{24}). These {\it
parametrized Minkowski theories} have first class constraints
(corresponding to the superhamiltonian and supermomentum ones of
ADM canonical gravity and resulting  from a deparametrization of
general relativity), which imply that the description is {\it
independent} from the choice of the 3+1 splitting. As a
consequence, the embeddings $z^{\mu}(\tau ,\vec \sigma )$ are the
{\it gauge variables of this special relativistic type of general
covariance}. Since, given the embeddings, the field of unit
normals to the hypersurfaces and the evolution vector
$\partial_{\tau}\, z^{\mu}(\tau ,\vec \sigma )$ give rise to two
(non-rotating the former, rotating the latter) congruences of
timelike observers, parametrized Minkowski theories are also
theories {\it describing timelike arbitrarily accelerated
observers}, where to learn how to study special relativistic
systems in {\it non-inertial} reference frames. Therefore, the
Shanmugadhsan canonical transformations and the canonical
reductions must be reformulated on arbitrary spacelike
hypersurfaces. In special relativity the foliations with spacelike
{\it hyperplanes} are particularly important: they correspond to
global accelerated reference frames generalizing the global
Galilei ones of the non-relativistic Newton mechanics. When we add
the gauge fixings, which restrict the hypersurfaces to
hyperplanes, the embedding gauge variables $z^{\mu}(\tau ,\vec
\sigma )$ are reduced to only ten: $z^{\mu}(\tau ,\vec \sigma ) =
x^{\mu}_s(\tau ) + b^{\mu}_r(\tau )\, \sigma^r$. While the
centroid $x^{\mu}_s(\tau )$ (with arbitrary velocity ${\dot
x}^{\mu}_s(\tau )$) describes the origin of the 3-coordinates
$\vec \sigma$ on each hyperplane, the spatial triad
$b^{\mu}_r(\tau )$ together with the unit normal $l^{\mu} =
b^{\mu}_{\tau}$ form an orthonormal tetrad. However the
$\tau$-independency of $l^{\mu}$ reduces to three the independent
degrees of freedom in $b^{\mu}_r(\tau )$ (three Euler angles
descring a rotating spatial reference frame). Only ten first class
constraints survive and $l^{\mu} = const$ is the gauge fixing for
three of them.

Moreover, for every timelike configuration of the isolated system
there is a foliation whose hyperplanes are intrinsically defined
by the configuration itself: the one with the hyperplanes
orthogonal to its conserved total 4-momentum. By definition this
foliation defines the {\it rest frame } of the configuration. To
select this foliation, the tetrad $b^{\mu}_A$ has to be gauge
fixed to coincide with the polarization vectors
$\epsilon^{\mu}_A(u(p_s))$ [$u^{\mu}(p_s) = p^{\mu}_s /
\sqrt{p^2_s}$], columns of the standard Wigner boost for timelike
Poincare' orbits (for this reason these hyperplanes are named {\it
Wigner hyperplanes}). After this fixation the only surviving
canonical variables are

i) a canonical but non-covariant variable ${\tilde x}^{\mu}_s$
replacing the centroid $x^{\mu}_s$ and the conjugate momentum
$p^{\mu}_s$, weakly equal to the total 4-momentum $P^{\mu}$ of the
system;

ii) canonical variables for the system living inside the Wigner
hyperplanes, which are either Lorentz scalars or Wigner-covariant
tensors.

Four first class constraints remain:

i) one, $\sqrt{p^2_s} - {\cal M} \approx 0$, identifies the
invariant mass of the isolated system as the mass associated with
${\tilde x}^{\mu}_s$;

ii) the other three, $\vec P \approx 0$, say that the Wigner
hyperplanes are the rest frame.

The variable ${\tilde x}^{\mu}_s$ is playing the role of a {\it
decoupled point particle clock}, since it describes the
relativistic {\it external} canonical non-covariant 4-center of
mass of the isolated system and it is the only variable which
breaks Lorentz invariance whichever is the system ({\it
universality of the breaking}). Associated with it there is an
{\it external} canonical realization of the Poincare' group, whose
Lorentz boost generators induce Wigner rotations on the Wigner
tensors living inside Wigner hyperplanes.

This {\it external} viewpoint describes the relativistic version
of the separation of the center of mass and defines the {\it
internal} canonical variables for the system inside the Wigner
hyperplanes. Due to the rest-frame constraints $\vec P \approx 0$
three of these degrees of freedom define a {\it internal gauge}
3-center of mass in each Wigner hyperplane. If we eliminate it
with three gauge fixings \footnote{It is convenient to force it to
coincide with the centroid $x^{\mu}_s$, origin of the
3-coordinates.}, we remain with only {\it internal relative
Wigner-covariant } canonical variables for the system. Then, if we
eliminate the last constraint by identifying the parameter $\tau$,
labeling the hyperplanes, with the rest-frame scalar time $T_s =
p_s \cdot {\tilde x}_s / \sqrt{p^2_s} = p_s \cdot x_s /
\sqrt{p^2_s}$, we obtain a new instant form of dynamics, the {\it
rest-frame instant form}, with a decoupled external 4-center of
mass and internal relative canonical variables. There is also a
{\it unfaithful internal} canonical realization of the Poincare'
group. This kinematical framework has allowed the definition of a
new kinematics for the relativistic N-body problem \cite{25}, a
study of relativistic rotational kinematics \footnote{Since
non-relativistic concepts like Jacobi coordinates, relative masses
and tensors of inertia are not extensible to the relativistic
level, a new treatment of rotational kinematics for deformables
bodies was needed. This has been found in Ref.\cite{26}, where
{\it dynamical body frames} and {\it canonical spin bases} were
defined and shown to be extensible to special relativity. } and of
multipole expansions \cite{27}. The extension of this kinematics
to relativistic perfect fluids and extended bodies is under
investigation \cite{28}.

Then we looked for a formulation of general relativity with matter
such that the switching off of the Newton constant $G$ would
produce the description of the same matter in the rest-frame
instant form of parametrized Minkowski theory with the general
relativistic general covariance deparametrizing to the special
relativistic one.

We started with the following family of non-compact spacetimes
\cite{1,24,29,30}:

i) globally hyperbolic, so that the ADM Hamiltonian formulation is
well defined if we start from the ADM action instead that from the
Hilbert one;

ii) topologically trivial, so that they can be foliated with
spacelike hypersurfaces diffeomorphic to $R^3$ (3+1 splitting of
spacetime with $\tau$, the scalar parameter labeling the leaves,
as a mathematical time);

iii) asymptotically flat at spatial infinity and with boundary
conditions at spatial infinity independent from the direction, so
that the {\it spi group} of asymptotic symmetries is reduced to
the Poincare' group with the ADM Poincare' charges as generators
\footnote{When we switch off the Newton constant $G$, the ADM
Poincare' charges in adapted coordinates become the generators of
the internal Poincare' group of parametrized Minkowski theories.}.
In this way we can eliminate the {\it supertranslations}, which
are the obstruction to define angular momentum in general
relativity, and we have the type of boundary conditions which are
needed to get well defined non-Abelian charges in Yang-Mills
theory, opening the possibility of a unified description of the
four interactions with all the fields belonging to same type of
function space. All these requirements imply that the {\it allowed
foliations} of spacetime must have the spacelike hyperplanes
tending in a direction-independent way to Minkowski spacelike
hyperplanes at spatial infinity, which moreover must be orthogonal
there to the ADM 4-momentum. But these are the conditions
satisfied by the singularity-free Christodoulou-Klainermann
spacetimes \cite{31}, in which the allowed hypersurfaces define
the {\it rest frame of the universe} and naturally become the
Wigner hyperplanes when $G \mapsto 0$. Therefore there are {\it
preferred asymptotic inertial observers}, which can be identified
with the {\it fixed stars}. These allowed hypersurfaces have been
called {\it Wigner-Sen-Witten hypersurfaces}, because it can be
shown that the Frauendiener reformulation \cite{32} of Sen-Witten
equations for triads allows (after the restriction to the
solutions of Einstein's equations) to transport the asymptotic
tetrads of the inertial observers in each point of the
hypersurface, generating a {\it local compass of inertia} to be
used to define {\it rotations with respect to the fixed stars}
\footnote{Instead the standard Fermi-Walker transport of the
tetrads of a timelike observer is a standard of non-rotation with
respect to a local oserver in free fall.} .

iv) All the fields have to belong to suitable weighted Sobolev
spaces so that the allowed spacelike hypersurfaces are Riemannian
3-manifolds without Killing vectors: in this way we avoid the
analogue of the Gribov ambiguity in general relativity and we can
get a unification of the function spaces of gravity and particle
physics.

After all these preliminaries it is possible to study the
Hamiltonian formulation of both ADM metric \cite{24} and tetrad
\cite{29,30} gravity \footnote{More natural for the coupling to
the fermions of the standard model of particles. Moreover tetrad
gravity is naturally a theory of timelike accelerated observers,
generalizing the ones of parametrized Minkowski theoris. In Refs.
\cite{29,30} there is a new parametrization of tetrad gravity,
still utilizing the ADM action, which emphasizes this aspect and
allows to solve 13 of the 14 constraints. } with their (8 and 14
respectively) first class constraints as generators of the
Hamiltonian gauge transformations. Then it is possible to look, at
least at a heuristic level, for Shanmugadhasan canonical
transformations performing the division between the gauge
variables and the Dirac observables for the gravitational field. A
complete exposition of these topics is in Refs.\cite{24,29,30},
where it is shown that it is possible to define a {\it rest frame
instant form of gravity} in which the effective Hamiltonian for
the evolution is the {\it ADM energy} $E_{ADM}$ \footnote{The
superhamiltonian constraint generates {\it normal} deformations of
the spacelike hypersurfaces, which are {\it not} interpreted as a
time evolution (like in the Wheeler-DeWitt approach) but as the
Hamiltonian gauge transformations ensuring that the description of
gravity is independent from the 3+1 splitting of spacetime like in
parametrized Minkowski theories.}.

Let us consider ADM canonical metric gravity. After an allowed 3+1
splitting of spacetime with spacelike hypersurfaces
$\Sigma_{\tau}$, the 4-metric ${}^4g_{AB}(\tau ,\vec \sigma )$ in
adapted coordinates is replaced with the lapse $N(\tau ,\vec
\sigma )$ and shift $N_r(\tau ,\vec \sigma )$ functions and with
the 3-metric ${}^3g_{rs}(\tau ,\vec \sigma )$ on $\Sigma_{\tau}$.
The conjugate momenta are $\pi_N(\tau ,\vec \sigma )$,
$\pi_N^r(\tau ,\vec \sigma )$, ${}^3\Pi^{rs}(\tau ,\vec \sigma )$,
respectively. There are four primary constraints  $\pi_N(\tau
,\vec \sigma ) \approx 0$, $\pi_N^r(\tau ,\vec \sigma ) \approx 0$
and four secondary ones: the superhamiltonian constraint ${\cal
H}(\tau ,\vec \sigma ) \approx 0$ and the supermomentum
constraints ${\cal H}^r(\tau ,\vec \sigma ) \approx 0$. Therefore
there are {\it eight arbitrary gauge variables}, four of which are
the lapse and shift functions. All the constraints are first class
and the Dirac Hamiltonian is \footnote{As shown in Ref.\cite{24}
the surface terms involve the ADM Poincare' charges.} $H_D = \int
d^3\sigma\, \Big( N(\tau ,\vec \sigma )\, {\cal H}(\tau ,\vec
\sigma ) + N_r(\tau ,\vec \sigma )\, {\cal H}^r(\tau ,\vec \sigma
) + \lambda_N(\tau ,\vec \sigma )\, \pi_N(\tau ,\vec \sigma ) +
\lambda_{N\, r}(\tau ,\vec \sigma )\, \pi^r_N(\tau ,\vec \sigma )
\Big) + (surface\, terms)$. The arbitrary functions
$\lambda_N(\tau ,\vec \sigma )$, $\lambda_{N\, r}(\tau ,\vec
\sigma )$ are the Dirac multipliers, which are the source of the
indeterminism in the Hamilton equations. The Hamiltonian version
\cite{33} of the {\it hole argument} amounts to fix completely the
gauge freedom of the Dirac multipliers and of the lapse and shift
functions outside the hole ${\cal A}$, but leaving them arbitrary
inside the hole.

As shown in Ref.\cite{24} the correct procedure to add the gauge
fixings is the following:

i) add three gauge fixings $\chi_r(\tau ,\vec \sigma ) \approx 0$
to the secondary supermomentum constraints: this amounts to a
choice of 3-coordinates on $\Sigma_{\tau}$. The requirement of
time constancy of the constraints $\chi_r(\tau ,\vec \sigma )
\approx 0$ will generate three gauge fixings $\varphi_r(\tau ,\vec
\sigma ) \approx 0$ for the primary constraints $\pi^r_N(\tau
,\vec \sigma ) \approx 0$, which determine the shift functions
$N_r(\tau ,\vec \sigma )$ (and therefore the gravitomagnetic
aspects and the eventual anisotropy of light propagation). The
time constancy of the $\varphi_r$'s will determine the Dirac
multipliers $\lambda_{N\, r}$'s.

ii) add a gauge fixing $\chi (\tau ,\vec \sigma ) \approx 0$ to
the secondary superhamiltonian constraint, which determines the
form of the spacelike hypersurface $\Sigma_{\tau}$ (it is a
statement about its extrinsic curvature). Its time constancy
produces the gauge fixing $\varphi (\tau ,\vec \sigma ) \approx 0$
for the primary constraint $\pi_N(\tau ,\vec \sigma ) \approx 0$,
which determines the lapse function $N(\tau ,\vec \sigma )$,
i.e.how the surfaces $\Sigma_{\tau}$ are packed in the foliation.
Now the 3+1 splitting of spacetime is completely determined and
the time constancy of $\varphi (\tau ,\vec \sigma ) \approx 0$
determines the Dirac multiplier $\lambda_N(\tau ,\vec \sigma )$. A
posteriori after having solved the Hamilton equations one could
find the embedding $z^{\mu}(\tau ,\vec \sigma )$ of these
Wigner-Sen-Witten hypersurfaces into the spacetime.

At this stage the canonical reduction is completed by going to
Dirac brackets, the Dirac Hamiltonian reduces to the surface term,
which can be shown \cite{24} to be equivalent to the ADM energy.
Therefore it becomes the effective Hamiltonian for the gauge
invariant observables parametrizing the reduced phase space.

To find a canonical basis of Dirac observables for the
gravitational field in absence of known solutions of the
superhamiltonian constraint, we can perform a quasi-Shanmugadhasan
canonical transformation adapted to only seven of the constraints
and utilize the information (see Ref.\cite{24} for its
justification) that this constraint has to be interpreted as the
{\it Lichnerowicz equation} for the conformal factor $\phi (\tau
,\vec \sigma ) = (det\, {}^3g(\tau ,\vec \sigma ) )^{1/ 12} =
e^{q(\tau ,\vec \sigma )/2}$ of the 3-metric. The result is ($a =
1,2$)

\begin{equation}
\begin{minipage}[t]{3cm}
\begin{tabular}{|l|l|l|} \hline
$N$ & $N_r$ & ${}^3g_{rs}$ \\ \hline $\pi_N \approx 0$ & $ \pi_N^r
\approx 0 $ & ${}^3{\tilde \Pi}^{rs}$ \\ \hline
\end{tabular}
\end{minipage} \hspace{1cm} {\longrightarrow \hspace{.2cm}} \
\begin{minipage}[t]{4 cm}
\begin{tabular}{|ll|l|l|l|} \hline
$N$ & $N_r$ & $\xi^{r}$ & $\phi$ & $r_{\bar a}$\\ \hline
 $\approx 0$ & $\approx 0$
& $\approx 0$ &
 $\pi_{\phi}$ & $\pi_{\bar a}$ \\ \hline
\end{tabular}
\end{minipage}.
\end{equation}

\noindent where $\xi^r(\tau ,\vec \sigma )$ are a parametrization
of the group manifold of the passive 3-diffeomorphisms of
$\Sigma_{\tau}$, describing its changes of 3-coordinates. The
Hamiltonian {\it gauge variables} are the seven configuration
variables $N(\tau ,\vec \sigma )$, $N_r(\tau ,\vec \sigma )$,
$\xi^r(\tau ,\vec \sigma )$ (they depend on the 4-metric and its
space gradients) and the momentum $\pi_{\phi}(\tau ,\vec \sigma )$
conjugate to the conformal factor (it depends also on the time
derivative of the 4-metric). The variables $\xi^r(\tau ,\vec
\sigma )$ and $\pi_{\phi}(\tau ,\vec \sigma )$ can be thought as a
possible 4-coordinate system with the Lorentz signature given by
the pattern "3 configuration + 1 momentum" variables.

The physical deterministic degrees of freedom of the gravitational
field are the non-local \footnote{This non-locality can be
considered a manifestation of {\it Mach's principle}: the
knowledge of the full 3-space at each time is needed to determine
the physics in the local neighborhood of each point of spacetime.}
Dirac observables (their expression in terms of the original
variables is not known) $r_a(\tau ,\vec \sigma )$, $\pi_a(\tau
,\vec \sigma )$, $a=1,2$, which in general are not Bergmann
observables being non-tensorial and coordinate-dependent. Even if
we do not know the solution $\phi = \phi [\xi^r, \pi_{\phi}, r_a,
\pi_a]$ of the Lichnerowicz equation, the class of Hamiltonian
gauges defined by the gauge fixing $\chi (\tau ,\vec \sigma ) =
\pi_{\phi}(\tau ,\vec \sigma ) \approx 0$ has the special property
that the Dirac observables $r_a(\tau ,\vec \sigma )$, $\pi_a(\tau
,\vec \sigma )$ remain canonical also at the level of Dirac
brackets. It is possible to study how to solve all the other
constraints (also in tetrad gravity \cite{30}) and how to express
all the original variables in terms of the Dirac observables
associated to a chosen gauge. This allows for the first time to
arrive at a completely fixed Hamiltonian gauge of the 3-orthogonal
type \footnote{Namely with ${}^3g_{rs}(\tau ,\vec \sigma )$
diagonal and with ${}^3g_{rr}(\tau ,\vec \sigma ) = f_r(r_a(\tau
,\vec \sigma ))$. The 3-orthogonal class of gauges seems to be the
nearest one to the physical laboratories on the Earth. Let us
remember that the standards of length and time are coordinate
units and not Bergmann observables \cite{34}. }, which when
restricted to the solutions of Einstein's equations (i.e. {\it
on-shell}) is equivalent to a well defined choice of 4-coordinates
for spacetime. It is now under investigation how to find a
post-Minkowskian approximation to Einstein spacetimes based on the
linearization of the theory in this completely fixed Hamiltonian
gauge so to make contact with the theory of gravitational waves.

It is evident that the Hamiltonian gauge variables of canonical
gravity carry an information about observers in spacetime, so that
they are not inessential variables like in electromagnetism and
Yang-Mills theory but take into account the fact that in general
relativity global inertial reference frames do not exist
\footnote{The equivalence principle only allows the existence of
local inertial frames along timelike geodesics describing the
worldline of a scalar test particle in free fall. }.

The separation between gauge variables and Dirac observables is an
extra piece of (non-local) information \cite{10,33}, which has to
be added to the equivalence principle, asserting the local
impossibility to distinguish gravitational from inertial effects,
to visualize which of the local forces acting on test matter are
{\it generalized inertial (or fictitious) forces depending on the
Hamiltonian gauge variables} and which are {\it genuine
gravitational forces depending on the Dirac observables, which are
absent in Newtonian gravity} \footnote{When we will introduce
dynamical matter, this Hamiltonian procedure will lead to
distinguish among action-at-a-distance, gravitational and apparent
effects. It will be important to see the implications on concepts
like gravitational passive and active masses and more in general
on the problem of the origin of inertia and its connection with
the various formulations of the Mach principle.}. Both types of
forces have a {\it different appearance in different 4-coordinate
systems}. In every 4-coordinate system (on-shell completely fixed
Hamiltonian gauge)

i) the {\it genuine tidal gravitational forces} in the geodesic
deviation equation will be well defined gauge-dependent
functionals only of the Dirac observables associated to that
gauge, so that Dirac's observables can be considered {\it
generalized non-local tidal degrees of freedom} ;

ii) the geodesics will have a different geometrical form which
again is functionally dependent on the Dirac observables in that
gauge;

iii) the description of the relative 3-acceleration of a free
particle in free fall given in the local rest frame of an observer
will generated various terms identifiable with the general
relativistic extension of the non-relativistic inertial
accelerations and again these terms will depend on both the Dirac
observables and the Hamiltonian gauge variables \footnote{See the
local interpretation \cite{35} of {\it inertial} forces as effects
depending on the {\it choice of a congruence of time-like
observers with their associated tetrad fields} as a reference
standard for their description. In metric gravity these tetrad
fields are used only to rebuild the 4-metric. The real theory
taking into account all the properties of the tetrad fields is
{\it tetrad gravity}. Note that the definition of
gravito-magnetism as the effects induced by ${}^4g_{\tau r}$ is a
pure inertial effect, because it is determined by the shift gauge
variables.}.

Therefore the Hamiltonian gauge variables, which change value from
a gauge to another one, {\it describe the change in the
appearance} of both the physical and apparent gravitational forces
going (on-shell) from a coordinate system to another one. This is
similar to what happens with non-relativistic inertial forces,
which however describe only apparent effects due to the absence of
genuine dynamical degrees of freedom (Dirac observables) in
Newtonian gravity. At the non-relativistic level, Newtonian
gravity is described only by action-at-a-distance forces and, in
absence of matter, there are {\it no tidal} forces among test
particles, since they are determined by the variation of the
action-at-a-distance force on the test particle created by the
Newton potential of a massive body. Instead in vacuum general
relativity the geodesic deviation equation shows that tidal
forces, locally described by the Riemann tensor, are acting on
test particles also in absence of every kind of matter: on-shell,
in any chosen 4-coordinate system, these tidal forces are
functionals of the non-local Dirac observables of the
gravitational field in the completely fixed Hamiltonian gauge
which corresponds on-shell to the chosen 4-coordinates.
Independently of gravity, in Newtonian physics we speak of {\it
global inertial (or fictitious) forces proportional to the mass}
when we look at matter not from an inertial reference frame
\footnote{See Ref.\cite{36} for the determination of {\it
quasi-inertial reference frames} in astronomy as those frames in
which rotational and linear acceleration effects are under the
sensibility threshold of the measuring instruments.} but from an
accelerated Galilean reference frame \footnote{With arbitrary
global translational and rotational 3-accelerations}. If the
non-inertial reference frame has translational acceleration $\vec
w(t)$ and angular velocity $\vec \omega (t)$ with respect to a
given inertial frame, a particle with free motion ($\vec a =
{\ddot {\vec x}} = 0$) in the inertial frame has the following
acceleration in the non-inertial frame

\begin{equation}
 {\vec a}_{NI} = - \vec w(t) + \vec x \times {\dot {\vec
\omega}}(t) + 2 {\dot {\vec x}} \times \vec \omega (t) + \vec
\omega (t) \times [\vec x \times \vec \omega (t)].
 \label{2}
 \end{equation}

\noindent After multiplication of this equation by the particle
mass, the second term on the right hand side is the {\it Jacobi
force}, the third term the {\it Coriolis force} and the fourth one
the {\it centrifugal force}.

In Ref.{37} a description, generally covariant under arbitrary
passive Galilean coordinate transformations [$t^{'} = T(t)$,
${\vec x}^{'} = \vec f (t, \vec x)$], of a free particle was
given. The analogue of Eq.(\ref{2}) contains more general {\it
apparent } forces, which reduce to those in Eq.(\ref{2}) in
particular rigid coordinate systems.

While in Newtonian physics an absolute reference frame is an
imagined extension of a rigid body and a clock (with any
coordinate systems attached), in general relativity  we must
replace the rigid body either by a cloud of test particles in free
fall (geodesic congruence) or by a test fluid (non-geodesic
congruence for non-vanishing pressure).

Therefore in general relativity, where there are no global
inertial reference frames, we have to use either a single
accelerated time-like observer or a congruence of accelerated
time-like observers with an associated conventionally chosen
either tetrad or field of tetrads. Usually this is done by
introducing {\it test observers} which describe the phenomena from
their kinematical point of view without introducing any (either
action or reaction) dynamical effect on the system (gravitational
field plus dynamical matter).

In the case of a single test observer with his tetrad, see
Ref.\cite{33}, after the choice of the local Minkowskian system of
(Riemann-Gaussian) 4-coordinates where the line element becomes
$ds^2 = - \delta_{ij} dx^i dx^j + 2 \epsilon_{ijk} x^j
{{\omega^k}\over c} dx^o dx^i + (1 + {{2 \vec a \cdot \vec x}\over
{c^2}} (dx^o)^2)$ (the constants $\vec a$ and $\vec \omega$ are
constant functionals of the Dirac observables of the gravitational
field in this particular gauge) , the test observer describes a
nearby time-like geodesics $y^{\mu}(\lambda )$ ($\lambda$ is the
affine parameter or proper time) followed by a test particle in
free fall with the following spatial equation: ${{d^2 \vec y}\over
{(dy^o)^2}} = - \vec a - 2 \vec \omega \times {{d \vec y}\over
{dy^o}} + {2\over {c^2}} \Big( \vec a \cdot {{d\vec y}\over
{dy^o}}\Big)\, {{d\vec y}\over {dy^o}}$ . If the test observer is
in free fall (geodesic observer) we have $\vec a = 0$. If the
triad of the test observer is Fermi-Walker transported (standard
of non-rotation of the gyroscope) we have $\vec \omega = 0$.

Therefore the relative acceleration of the particle with respect
to the observer with this special system of coordinates (replacing
the global non-inertial non-relativistic reference frame) is
composed by the observer 3-acceleration plus a relativistic
correction and by a Coriolis acceleration. With other coordinate
systems, other terms would appear. These are the inertial effects
due to the Hamiltonian gauge variables.

In conclusion different on-shell Hamiltonian gauge fixings ,
corresponding to on-shell variations of the Hamiltonian gauge
variables, give rise to {\it different appearances} of the
physical forces as gauge-dependent functionals of the Dirac
observables in that gauge of the type $F(r_{\bar a}, \pi_{\bar
a})$ (like $\vec a$ and $\vec \omega$ in the previous example).
Newtonian gravity is recovered with a double limit:

i) Zero curvature limit, which is obtained by sending to zero the
Dirac observables. In this way we get Minkowski space-time (a
solution of Einstein's equations) with those systems of
coordinates which are compatible with Einstein's theory. As shown
in Refs.\cite{24,30} this implies the vanishing of the Cotton-York
3-conformal tensor, namely that the allowed 3+1 splittings of
Minkowski space-time compatible with Einstein's equations have the
{\it leaves 3-conformally flat} in absence of matter.

ii) The  ($c \rightarrow \infty$) limit.

This implies that these functionals  must be rewritten as the
limit for vanishing Dirac observables of series in $1/c$ :
$F_{newton} = lim_{c \rightarrow \infty}\,\, lim_{r_{\bar a},
\pi_{\bar a} \rightarrow 0}\,\,\Big( F_o + {1\over c} F_1 +
...\Big) = F_o {|}_{r_{\bar a} = \pi_{\bar a} = 0}$. $F_{newton}$,
which may be coordinate dependent, becomes the {\it Newtonian
inertial force} in the corresponding general Galilean coordinate
system.

Let us come back to the problem of the physical identification of
the mathematical points as point-events \cite{10,33}, namely as
physical events. As shown in Ref.\cite{39} there are 14
algebraically independent curvature scalars for $M^4$, which are
reduced to four when Einstein equations without matter are used.
Bergmann and Komar \cite{40} discovered that the four (curvature
scalars, i.e. Bergmann observables) eigenvalues $\lambda_i(\tau
,\vec \sigma )$, $i=1,..,4$, of the Weyl tensor do {\it not
depend} on the lapse and shift functions and can be used to define
systems of {\it intrinsic pseudo-4-coordinates}
$F^{[A]}(\lambda_i(\tau ,\vec \sigma ))$ (we put the index $A$
between square brackets to denote the scalar character of the
functions $F^{[A]}$ under spacetime diffeomorphisms), where the
$F^{[A]}$'s are arbitrary functions restricted by the condition
$det\, | \{ F^{[A]}, {\cal H}^B \} | \not= 0$, where ${\cal H}^B =
( {\cal H}; {\cal H}^r) \approx 0$ are the secondary constraints.
This means that an admissible system of four gauge fixing
constraints (determining the gauge variables $\xi^r$ and
$\pi_{\phi}$) is

\begin{equation}
\chi^A(\tau ,\vec \sigma ) = \sigma^A -
F^{[A]}(\lambda_i(\tau,\vec \sigma )) \approx 0.
 \label{3}
 \end{equation}

\noindent Clearly these conditions break completely general
covariance by identifying coordinates with scalar fields.

As already said, the time constancy of these gauge fixings
determines the lapse and shift functions and then the Dirac
multipliers, so that at the end on the solutions of Einstein's
equations we get a unique 4-coordinate system $\sigma^A$ for the
mathematical 4-manifold $M$. But in this completely fixed gauge
the Weyl scalars become gauge-dependent functions of the Dirac
observables in this gauge, $\lambda_i(\tau ,\vec \sigma ){|}_G =
f_i^{(G)}(r_a(\tau ,\vec \sigma ), \pi_a(\tau ,\vec \sigma ))$.
Therefore, every 4-coordinate system of the mathematical spacetime
may be identified by means of {\it four dynamical individuating
fields} (in the terminology of Stachel \cite{8}) which depend only
on the Dirac observables in that gauge. Mathematical points of
spacetime are transformed in physical point-events by means of the
four independent degrees of freedom of the gravitational field.

In a sense {\it point-events of spacetime and the vacuum
gravitational field are synonymous}.

Then {\it test matter} has to be used to measure the gravitational
field, following Ref.\cite{41} with a material reference fluid
employing the intrinsic pseudo-coordinates.  We have an axiomatic
theory of measurement employing test matter \cite{42}, but no real
theory based on {\it dynamical matter} due to the absence of
solutions describing spacetimes without symmetries and due to the
difficulties in using dynamical point particles in general
relativity since the ultraviolet divergencies on the worldlines
are much worse that in electrodynamics. The practice of the
laboratories on and around the Earth is to use post-Newtonian
corrections to Newtonian gravity and/or special relativistic
theory of measurement for the electromagnetic phenomena.
Astrometry \cite{34} looks for the materialization of a global
non-rotating, quasi-inertial reference frame in the form of a
fundamental catalogue of stellar positions and proper motions,
while physics in the solar system employs a barycentric
post-Newtonian reference frame. Therefore, regarding matter,
general relativity is essentially a dualistic theory. The
situation becomes worse in the attempts of quantization: the need
that the mass of test matter must be big to simulate a classical
measuring apparatus  contradicts its being test and not dynamical
matter. There is a conflict between the role of mass as the charge
of gravity with all the implications of the equivalence principle
and quantum theory: it is a complete mystery which is the genesis
of mass and why it seems not to be quantized.

Let us remark that at the level of Dirac brackets there is an
induced non-commutative structure added to spacetime by the
functions $F^{[A]}$.

This is the final theoretical answer to the problem raised by the
{\it hole argument}, which however employs a division between
generalized inertial (the gauge variables) and tidal (the Dirac
observables) effects in which both types of effects have a
coordinate-dependent appearance, i.e. they are not Bergmann
observables.

Bergmann and Komar \cite{39,43} also defined an {\it intrinsic
(tetradic-like) 4-metric}

\begin{equation}
 {}^4g_{[A] [B]}(F^{[E]} ) = {{\partial \sigma^C}\over
{\partial F^{[A]} }}\, {{\partial \sigma^D}\over {\partial F^{
[B]}}}\, {}^4g_{CD}(\sigma^E ),
 \end{equation}

\noindent whose ten components are Bergmann observables, but did
not develop its implications.

The concept of an intrinsic 4-metric together with the (at least
local) existence of the quasi-Shanmugadhasan canonical
transformation leads to the following {\it conjecture}: there
should exist a class of quasi-Shanmugadhasan canonical
transformations such that both the resulting Hamiltonian gauge
variables (the generalized inertial effects) and Dirac observables
(the generalized tidal effects) are {\it also} Bergmann
observables.

If the conjecture is true, there is a {\it preferred} family of
canonical bases in the ADM phase space worthy of investigation
both for trying to solve the superhamiltonian constraint and for a
new attempt to the canonical quantization of gravity, in which
only the physical (but Bergmann observable) degrees of freedom of
the gravitational field, and not the inertial effects, are
quantized, preserving in this way the causal structure of
spacetime.

Let us remark that neither string theory nor loop quantum gravity
have developed a strategy for finding the solution of the
superhamiltonian constraint. As a consequence, we do not know the
modifications of Newton law between two bodies at short distances
(but which ones at the classical level? Planck length is a quantum
effect depending on $\hbar$) induced by Einstein's general
relativity, since it depends on the conformal factor of the
3-metric which has to be found as the solution of Lichnerowicz
equation.

Let us conclude with a comment on the possibility to arrive at an
{\it operational definition of a region of spacetime} by means of
the technological developments connected with the Global
Positioning System (GPS) \cite{44}. In Ref.\cite{10} there is the
description of GPS and of a possible modified (i.e. not using the
gravitational field of the Earth as input) experimental setup and
protocol for positioning and orientation, which should allow a
physical individuation of point-events in regions with non weak
fields like near the Sun or Jupiter (see Refs.\cite{45} for other
proposals). One should employ a net of satellites to establish a
{\it radar-gauge system of 4-coordinates} $\sigma^A_{(R)} = (
\tau_{(R)}; \sigma^r_{(R)})$ in a finite region ($\tau_{(R)} =
const$ defines the radar simultaneity surfaces). Each satellite
may be thought as a timelike observer (the satellite worldline)
with a tetrad (the timelike vector is the satellite 4-velocity and
the spatial triad is built with gyroscopes) and one of them is
chosen as the origin of the radar-4-coordinates. Either by using
test polarized light to measure the relative spatial rotation of
triads or by measuring the motion of $n \geq 4$ test particles
\cite{46} it should be possible to measure the components of the
4-metric in these radar-gauge coordinates. Then it is a matter of
computation to check:

i) whether Einstein's equation in radar-gauge coordinates are
satisfied;

ii) which are the functions $F^{[A]}$ to be used in Eqs.(\ref{3})
to identify the radar-gauge by means of the intrinsic coordinate
method.

This procedure would close the {\it coordinate circuit} of general
relativity, linking individuation to experimentation \cite{10}.


\begin{thebibliography}{99}

\bibitem{1}L.Lusanna, {\it Towards a Unified Description of the Four
Interactions in Terms of Dirac-Bergmann Observables}, invited
contribution to the book {\it Quantum Field Theory: a 20th Century
Profile} of the Indian National Science Academy, ed. A.N.Mitra,
foreward F.J.Dyson (Hindustan Book Agency, New Delhi, 2000)
(HEP-TH/9907081).

\bibitem{2}G.Leibbrandt, {\it Status Quo of the Coulomb Gauge:
Problems and a Possible Remedy}, Nucl.Phys. (Proc.Suppl.) {\bf
B64}, 101 (1998).

\bibitem{3}S.Shabanov, {\it Geometry of the Physical Phase Space
in Quantum Gauge Systems}, Phys.Rep. {\bf 326}, 1 (2000).

\bibitem{4}L.Lusanna, {\it Classical Yang-Mills Theory with
Fermions: I. General Properties of a System with Constraints},
Int.J.Mod.Phys. {\bf A10}, 3531 (1995). {\it Classical Yang-Mills
Theory with Fermions: II. Dirac's Obsevables }, Int.J.Mod.Phys.
{\bf A10}, 3675 (1995).


\bibitem{5}M.Asorey, {\it Maximal Non-Abelian Gauges and Topology
of Gauge Orbit Space}, e-print 1998 (hep-th/9812174).

\bibitem{6}L. Lusanna, {\it The N- and 1-Time Classical Descriptions of N-Body
Relativistic Kinematics and the Electromagnetic Interaction}, Int.
J. Mod. Phys. {\bf A12}, 645 (1997).

\bibitem{7}A.Einstein, {\it Die formale Grundlage der allgemeinen
Relativit\"atstheorie}, in Preuss.Akad. der Wiss.Sitz. 1030
(1914). {\it Die Grundlage der allgemeinen Relativit\"atstheorie},
Annalen der Physik {\bf 49}, 769 (1916) (translation by W.Perrett
and G.B.Jeffrey, {\it The Foundations of the General Theory of
Relativity}, in {\it The Principle of Relativity} (Dover, New
York, 1952)).


\bibitem{8}J.Stachel, {\it Einstein's Search for General
Covariance, 1912-1915}, in {it Einstein and the History of General
Relativity}, Einstein studies, Vol. 1, eds. D.Howard and J.Stachel
(Birkh\"auser, Boston, 1986). {\it The Meaning of General
Covariance - The Hole Story}, in {\it Philosophical Problems of
the Internal and External Worlds, Essays on the Philosophy of
Adolf Gruenbaum}, eds. J.Earman, I.Janis, G.J.Massey and N.Rescher
(University of Pittsburgh Press, Pittsburgh, 1993).

\bibitem{9}J.Norton, {\it How Einstein found his Field Equations:
1912-1915}, Historical Studies in the Physical Sciences {\bf 14},
252 (1984). {\it Einstein, the Hole Argument and the Reality of
Space}, in {\it Measurement, Realism and Objectivity}, ed. J.Forge
(Reidel, Dordrecht, 1987). {\it The Hole Argument}, PSA 1988, Vol.
2. {\it Coordinate and Covariance: Einstein's View of Space-Time
and the Modern View}, Found.Phys. {\bf 19}, 1215 (1989). {\it The
Physical Content of General Covariance}, in {\it Studies in the
History of General Relativity}, Einstein Studies, Vol. 3, eds.
J.Eisenstaedt and A.Kox (Birkh\"auser, Boston, 1992). {\it General
Covariance and the Foundations of General Relativity: Eight
Decades of Dispute}, Rep.Prog.Phys. {\bf 56}, 791 (1993). {\it
Einstein;s Triumph over the Space-Time Coordinate System}, talk at
the Conference in honor of R.Torretti, Univ. of Puerto Rico 2001
(e-print PITT-PHIL-SCI00000380).

\bibitem{10} M.Pauri and M.Vallisnari, {\it Ephemeral Point-Events: is there a
Last Remnant of Physical Objectivity?}, essay for the 70th
birthday of R.Torretti, 2002(gr-qc/0203014).

\bibitem{11}J.Earman and J.Norton, {\it What Price Spacetime
Substantivalism? The Hole Story}, British J. for the Philosophyof
Science {\bf 38}, 515 (1987).

\bibitem{12}R.Rynasiewicz, {\it Absolute versus Relational
Space-Time: An Outmoded Debate?}, Journal of Philosophy {\bf 43},
279 (1996).

\bibitem{13}P.G.Bergmann and A.Komar, {\it The Coordinate Group
Symmetries of General Relativity}, Int.J.Theor.Phys. {\bf 5}, 15
(1972).

\bibitem{14}P.G.Bergmann, {\it Observables in General Relativity},
Rev.Mod.Phys. {\bf 33}, 510 (1960).

\bibitem{15}C.Rovelli, {\it What is Observable in Classical and
Quantum Gravity?}, Class. Quantum Grav. {\bf 8}, 297
(1991).\hfill\break
 J.D.Brown and K.Kuchar, {\it Dust as a Standard of Space and Time
 in Canonical Quantum Gravity}, Phys.Rev. {\bf D51}, 5600
 (1995).\hfill\break
 J.D.Brown and D.Marolf, {\it Relativistic Material Referene
 Systems}, Phys.Rev. {\bf D53}, 1835 (1996).

\bibitem{16}M.Pauri, {\it Realta' e Oggettivita'}, in {\it
L'Oggettivita' nella Conoscenza Scientifica} (F.Angeli, Brescia,
1996).

\bibitem{17}S.Shanmugadhasan, {\it Canonical Formalism for
Degenerate Lagrangians}, J.Math.Phys. {\bf 14},
677 (1973).\hfill\break
 L.Lusanna, {\it The Shanmugdhasan Canonical Transformation,
 Function Groups and the Extended Second Noether Theorem},
 Int.J.Mod.Phys. {\bf A8}, 4193 (1993).


\bibitem{18}P.A.M. Dirac, {\it Lectures on Quantum Mechanics},
Belfer Graduate School of Science (Yeshiva University, New York,
N.Y., 1964).

\bibitem{19}L.Lusanna, {\it An Enlarged Phase Space for Finite-Dimensional
Constrained Systems, Unifying their Lagrangian, Phase- and
Velocity-Space Descriptions}, Phys.Rep. {\bf 185}, 1 (1990). {\it
The Second Noether Theorem as the Basis of the Theory of Singular
Lagrangians and Hamiltonian Constraints }, Riv.Nuovo Cimento {\bf
14}, n.3, 1 (1991). {\it On the BRS's }, J.Math.Phys. {\bf 31},
428 (1990). {\it Lagranian and Hamiltonian Many-Time Equations},
J.Math.Phys. {\bf 31}, 2126 (1990). {\it Classical Observables of
Gauge Theories from the Multitemporal Approach}, Comtemp.Math.
{\bf 132}, 531 (1992).\hfill\break
 M.Chaichian, D.Louis Martinez and L.Lusanna, {\it Dirac's Constrained
 Systems: The Classification of Second-Class Constraints},
 Ann.Phys.(N.Y.) {\bf 232}, 40 (1994).

\bibitem{20}D. Alba and L. Lusanna, {\it The Lienard-Wiechert Potential of Charged
Scalar Particles and Their Relation to Scalar Elctrodynamics in
the Rest-Frame Instant Form}, Int. J. Mod. Phys. {\bf A13}, 2791
(1998).

\bibitem{21}H. Crater and L. Lusanna,  {\it The Rest-Frame Darwin Potential
from the Lienard-Wiechert Solution in the Radiation Gauge}, Ann.
Phys. {\bf 289}, 87 (2001) (hep-th/0001046).

\bibitem{22} D. Alba, H. Crater and L. Lusanna,  {\it The Semiclassical
Relativistic Darwin Potential for Spinning Particles in the
Rest-Frame Instant Form: Two-Body Bound States with Spin 1/2
Constituents}, Int. J. Mod.Phys. {\bf A16}, 3365 (2001)
(hep-th/0103109).

\bibitem{23} D. Alba and L. Lusanna, {\it The Classical Relativistic Quark Model
in the Rest-Frame Wigner-Covariant Coulomb Gauge}, Int. J. Mod.
Phys. {\bf A13}, 3275 (1998).

\bibitem{24}L.Lusanna, {\it The Rest-Frame Instant Form of Metric Gravity},
Gen.Rel.Grav. {\bf 33}, 1579 (2001)(gr-qc/0101048).

\bibitem{25}D. Alba, L. Lusanna and M. Pauri, {\it Centers of Mass and
Rotational Kinematics for the Relativistic N-Body Problem in the
Rest-Frame Instant Form}, J. Math. Phys. {\bf 43}, 1677 (2002)
(hep-th/0102087).

\bibitem{26}D. Alba, L. Lusanna and M. Pauri, {\it Dynamical Body Frames,
Orientation-Shape Variables and Canonical Spin Bases for the
Non-Relativistic N-Body Problem}, J. Math. Phys. {\bf 43}, 373
(2002) (hep-th/0011014).


\bibitem{27}D. Alba, L. Lusanna and M. Pauri,{\it Multipolar Expansions
for the Relativistic N-Body Problem in the Rest-Frame Instant
Form} (hep-th/0103092).

\bibitem{28}L.Lusanna and D.Nowak-Szczepaniak, {\it The Rest-Frame Instant
Form of Relativistic Perfect Fluids with Equation of State $\rho =
rho (n,s)$ and of Non-Dissipative Elastic Materials},
Int.J.Mod.Phys. {\bf A15}, 4943 (2000) (hep=th/0003095).
\hfill\break
 D.Alba and L.Lusanna, in preparation.


\bibitem{29}L.Lusanna and S.Russo, {\it A New Parametrization for Tetrad Gravity},
Gen.Rel.Grav. {\bf 34}, 189 (2002)(gr-qc/0102074).

\bibitem{30}R.De Pietri, L.Lusanna, L.Martucci and S.Russo, {\it Dirac's
Observables for the Rest-Frame Instant Form of Tetrad Gravity in a
Completely Fixed 3-Orthogonal Gauge}, to appear in Gen.Rel.Grav.
(gr-qc/0105084).

\bibitem{31}D.Christodoulou and S.Klainermann, {\it The Global
Nonlinear Stability of the Minkowski Space} (Princeton, Princeton,
1993).

\bibitem{32}J.Frauendiener, {\it Triads and the Witten Equation},
Class. Quantum Grav. {\bf 8}, 1881 (1991).


\bibitem{33}L.Lusanna and M.Pauri, {\it General Covariance and the
Objectivity of  Space-Time Point-Events: the Role of Gravitational
and Gauge Degrees of Freedom in General Relativity}, in
preparation.

\bibitem{34}M.H.Soffel, {\it Relativity in Astrometry, Celestial
Mechanics and Geodesy} (Springer, Berlin, 1989).

\bibitem{35}R.J.Jantzen, P.Carini and D.Bini, {\it  The Many Faces
of Gravitoelectromagnetism}, Ann.Phys.(N.Y.) {\bf 215}, 1 (1992)
(gr-qc/0106043). {\it The Inertial Forces - Test Particle Motion
Game}, e-print 1998 (gr-qc/9710051).

\bibitem{36}J.Kovaleski, I.I.Mueller and B.Kolaczek, {\it
Reference Frames in Astronomy and Geophysics} (Kluwer, Dordrecht,
1989).

\bibitem{37}R.De Pietri, L.Lusanna and M.Pauri, {\it Standard
and Generalized Newton Gravities as 'Gauge' Theories of the
Extended Galilei Group: I. The Standard Theory}, Class. Quantum
Grav. {\bf 12}, 219 (1995).

\bibitem{38}C.W.Misner, K.S.Thorne and J.A.Wheeler, {\it
Gravitation} (Freeman, New York, 1973).\hfill\break
 H.Stephani, {\it General Relativity} (Cambridge University Press,
 Cambridge, 1996).

\bibitem{39}J.Ge'he'niau and R.Debever, {\it Les quatorze
invariants de courbure de l'espace riemannien a' quatre
dimensions}, in Proceedings of the {\it Jubilee of Relativity
Theory}, Bern 1955, eds. A.Mercier and M.Kervaire (Birkh\"auser,
Basel, 1956).

\bibitem{40}P.G.Bergmann and A.Komar, {\it Poisson Brackets
between Locally Defined Observables in General Relativity},
Phys.Rev.Lett. {\bf 4}, 432 (1960).



\bibitem{41}B.De Witt, {\it The Quantization of Geometry}, in {\it
Gravitation: An Introduction to Current Reasearch}, ed. L.Witten
(Wiley, New York, 1962).

\bibitem{42}J.Ehelers, F.A.E.Pirani and A.Schild, {\it The
Geometry of Free Fall and Light Propagation}, in {\it General
Relativity}, ed. L.O'Raifeartaigh (Clarendon, Oxford, 1972).

\bibitem{43}P.G.Bergmann, {\it The General Theory of Relativity},
in {\it Handbuch der Physik, Vol.IV, Principles of Electrodynamics
and Relativity}, ed. S.Flugge (Springer, Berlin, 1962).



\bibitem{44}N.Ashby and J.J.Spilker, {\it Introduction to
Relativistic Effects on the Global Positioning System}, in {\it
Global Positioning System: Theory and Applications}, Vol.1, eds.
B.W.Parkinson and J.J.Spilker (American Institute os Aeronautics
and Astronautics, 1995).

\bibitem{45}C.Rovelli, {\it GPS Observables in General
Relativity}, e-print 2001 (gr-qc/0110003).\hfill\break
 M.Blagojevic', J.Garecki, F.W.Hehl and Yu.N.Obukhov, {\it Real
 Null Coframes in General Relativity and GPS Type Coordinates},
 e-print 2001 (gr-qc/0110078).

\bibitem{46}I.Ciufolini and J.A.Wheeler, {\it Gravitation and
Inertia} (Princeton University Press, Princeton, 1995) pp.34-36.





\end{thebibliography}
\end{document}